\documentclass[preprint,3p,12pt,a4paper]{elsarticle}

\usepackage{graphicx,xcolor}

\biboptions{square,sort&compress,numbers}

\newcommand{\uu}[1]{\ensuremath{\,\mbox{#1}}}
\newcommand{\uuu}[1]{\ensuremath{\,[\mbox{#1}]}}

\newcommand{\corr}[1]{#1}

\journal{Scripta Materialia}

\begin{document}

\begin{frontmatter}


\title{Surface energies of AlN allotropes from first principles}
\author{David Holec\corref{cor1}}
\ead{david.holec@unileoben.ac.at}
\author{Paul H. Mayrhofer}
\cortext[cor1]{To whom all correspondence should be addressed.}
\address{Montanuniversi\"at Leoben, Franz-Josef-Stra\ss{e} 18, Leoben A-8700, Austria}
\address{\corr{Christian Doppler Laboratory for Application Oriented Coating Development at the Department of Physical Metallurgy and Materials Testing, Montanuniversität Leoben, Franz-Josef-Stra\ss{e} 18, Leoben A-8700, Austria}}

\begin{abstract}
In this Letter we present first principle calculation of surface energies of rock-salt (B1), zinc-blende (B3), and wurtzite (B4) AlN allotropes. Out of several low-index facets, the highest energies are obtained for mono-atomic surfaces (i.e. only by either Al or N atoms): $\gamma_{\{111\}}^{\rm B1}=410\uu{meV/\AA}^2$, $\gamma_{\{100\}}^{\rm B3}=346\uu{meV/\AA}^2$, $\gamma_{\{111\}}^{\rm B3}=360\uu{meV/\AA}^2$, and $\gamma_{\{0001\}}^{\rm B4}=365\uu{meV/\AA}^2$. The difference between Al- and N-terminated surfaces in these cases is less then $20\uu{meV/\AA}^2$. The stoichiometric facets have energies lower by $100\uu{meV/\AA}^2$ or more. The obtained trends could be rationalised by a simple nearest-neighbour broken-bond model.

\end{abstract}

\begin{keyword}
Aluminium nitride \sep surface energy \sep Density Functional Theory


\end{keyword}

\end{frontmatter}


\section{Introduction}
Aluminium nitride is a material with a fascinating variety of applications spanning from optoelectronic \cite{Jain2000a} and acoustic \cite{Turner1994} devices to the beneficial influence on mechanical properties (e.g., increased hardness) and performance of protective hard coatings \cite{Mayrhofer2003}. It crystallises in the wurtzite structure (B4, space group $P6_3mc$). Under non-equilibrium deposition parameters, its metastable zinc-blende variant (B3, space group $F4\bar3m$) can be stabilised \cite{Petrov1992}. Finally, under pressures above $\approx16\uu{GPa}$ \cite{Holec2010} it transforms to the rock-salt structure (B1, space group $Fm\bar3m$).

\corr{The knowledge of surface energies and the corresponding trends is important in several areas. As discussed by Gall \textit{et al.} \cite{Gall2003} for the case of TiN, surface energy is a decisive parameter for the thin film microstructure when grown under near-to-equilibrium conditions, i.e. when thermodynamics rather than kinetics control the texture formation. This is the case, for example, of the chemical vapour deposition of AlN for optoelectronic applications. Equally important is this quantity for the discussion of decomposition pathways of, e.g., unstable Ti$_{1-x}$Al$_x$N \cite{Mayrhofer2007,Rachbauer2011a}. Another emerging area of interest is the stabilisation of rock-salt AlN using a multilayer architecture for Ti$_{1-x}$Al$_x$N/AlN or Cr$_{1-x}$Al$_x$N/AlN coatings \cite{Chawla2012}, as it strengthens the material, and has the potential for stopping and/or deflecting cracks \cite{Schlogl2012}. Due to the nanostructured character of these composites, surface energy represents a non-negligible contribution for the overall energetic balance \cite{chawla_sub}.}

Although there exist plenty of papers, both on calculations as well as on experimental results, a systematic study of the surface energies is missing. In this Letter we use quantum mechanical calculations to obtain energies of several low-index clean surfaces (i.e., no surface reconstruction is considered here) of the B1, B3, and B4 allotropes. Since we cover several surface orientations as well as crystallographic structures, these results represent a coherent data set which can \corr{be directly used as is in above described applications.}

\section{Calculational details}
Quantum mechanical calculations within the framework of Density Functional Theory (DFT) were performed using the Vienna Ab initio Simulation Package (VASP) \cite{Kresse1996}.The projector augmented wave method pseudopotentials used Generalised Gradient Approximation (GGA) \cite{Kresse1999} as parametrised by Perdew and Wang \cite{Wang1991}. The first Brillouin zone (1BZ) of the bulk material \corr{unit cells} was meshed with $5\times5\times5$, $8\times8\times8$, and $9\times9\times6$ $k$-points distributed according to the Monkhorst-Pack scheme for the B1, B3, and B4 allotrope, respectively. \corr{The sampling was adopted for the surface supercells (see below), with only 1 $k$-point along the direction perpendicular to the surface.} The total energy was obtained by integration over the whole 1BZ using the tetrahedron method. These parameters together with plane wave cut-off energy of $450\uu{eV}$ ensure accuracy of the total energy of $\approx 1\uu{meV/atom}$.

For each structure and each surface orientation, an $N_{\rm cell}$-atom primitive cell was constructed such that the particular facet was perpendicular to the $z$-axis. Subsequently, $n$ primitive cells were stacked along the $z$-axis followed $h_{\rm vac}$ thick vacuum. Periodic boundary conditions were applied on those slabs. If the surface area is $A$ and contains $N_{\rm surf}$ atoms, the surface energy is calculated as
\begin{equation}
  \gamma(n, h_{\rm vac}) = \frac1{2A}(E_{\rm slab}-E_{\rm bulk}(n N_{\rm cell}+N_{\rm surf}))
\end{equation}
where $E_{\rm slab}$ is the total energy of the slab while $E_{\rm bulk}$ is the bulk total energy per atom of the respective structure. Positions of all atoms were relaxed during calculations of $E_{\rm slab}$.

Finally, $\gamma(n, h_{\rm vac})$ is converged simultaneously with respect to the number of primitive cells in the slab, $n$, and the vacuum thickness, $h_{\rm vac}$, in order to rule out undesired surface interactions either though the bulk material or though the separating vacuum.

\section{Results and discussion}

The calculated bulk lattice parameters (see Tab.~\ref{tab:alat}) agree well with both calculated and experimental values previously published. The lowest energy of formation is obtained for the wurztite B4 structure, closely followed by the zinc-blende B3 ($\Delta E_f^{\rm B3-B4}=0.0213\uu{eV/at.}$). The rock-salt structure B1 possesses highest energy of formation ($\Delta E_f^{\rm B1-B4}=0.1813\uu{eV/at.}$). These values correspond well with data calculated by Siegel \textit{et al. }\cite{Siegel2006}.

\begin{table}[t]
  \centering
  \caption{Lattice constants, $a$ and $c$, total energy, $E_{\rm tot}$, and energy of formation, $E_f$, of individual AlN structures. Calculated and experimental lattice parameters from the literature are given for comparison.}\label{tab:alat}
  \begin{tabular}{c|cccc|cccc}
       & \multicolumn{4}{c|}{this study} & \multicolumn{4}{c}{previous works}\\
       & $E_{\rm tot} \uuu{eV/at.}$ & $E_{f} \uuu{eV/at.}$ & $a\uuu{\AA}$  & $c\uuu{\AA}$ & \multicolumn{2}{c}{$a\uuu{\AA}$} & \multicolumn{2}{c}{$c\uuu{\AA}$}\\
      & & & & & calc. & expt. & calc. & expt. \\ \hline
    B1 & $-7.3202$ & $-1.3038$ & $4.070$ &         & $4.06$ \cite{Siegel2006} & $4.05$ \cite{JCPDF} & & \\
    B3 & $-7.4801$ & $-1.4637$ & $4.401$ &         & $4.39$ \cite{Siegel2006} & $4.38$ \cite{Petrov1992}& &\\
    B4 & $-7.5014$ & $-1.4850$ & $3.129$ & $5.016$ & $3.12$ \cite{Siegel2006} & $3.111$ \cite{Madelung2004} & $5.00$ \cite{Siegel2006} & $4.979$ \cite{Madelung2004}
  \end{tabular}
\end{table}

In this study we consider 4 low-index surfaces ($\{100\}$, $\{110\}$, $\{111\}$, and $\{112\}$) for the cubic B1 and B3 structures, and 3 different orientations ($\{0001\}$, $\{1\bar100\}$, and $\{11\bar20\}$) for the hexagonal B4 phase. The respective primitive cells contain 4, 4, 6, and 12 atoms for cubic structures and 4, 8, and 8 for the wurztite one. Their exact description is given in \ref{appendix}. It is worth noting that the B1-$\{111\}$, B3-$\{100\}$, B3-$\{111\}$, and B4-$\{0001\}$ planes contain only atoms of one specie. Consequently, in these cases we have calculated separately Al- and N-terminated surfaces.

\begin{figure}[!t]
  \centering
  \includegraphics{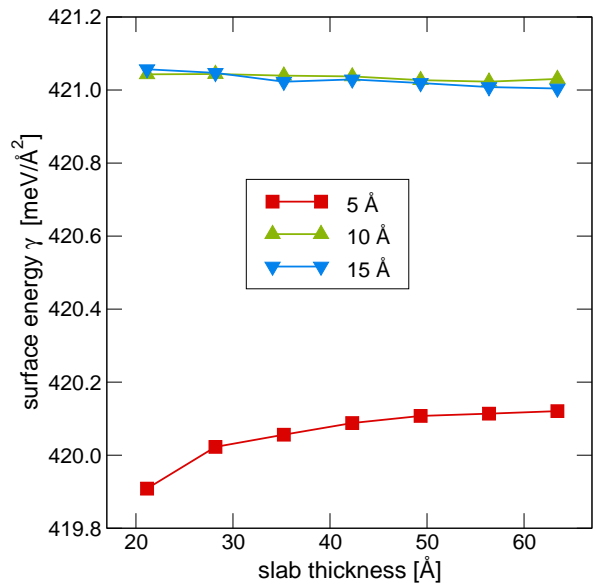}
  \caption{Convergence of the Al-terminated $\{111\}$ surface energy of B1-AlN.}\label{fig:conv}
\end{figure}

A typical example of the surface energy as a function of the slab and vacuum thickness is shown in Fig.~\ref{fig:conv} for the Al-terminated B1-$\{111\}$ surface. As in all other cases, the difference between results for $10\uu{\AA}$ and $15\uu{\AA}$ is negligible, while $5\uu{\AA}$ gives different $\gamma$ values. The slab thickness of $\approx20\uu{\AA}$ is sufficient in this case, but for other orientations thicker slabs are needed.

\begin{figure}[!t]
  \centering
  \includegraphics{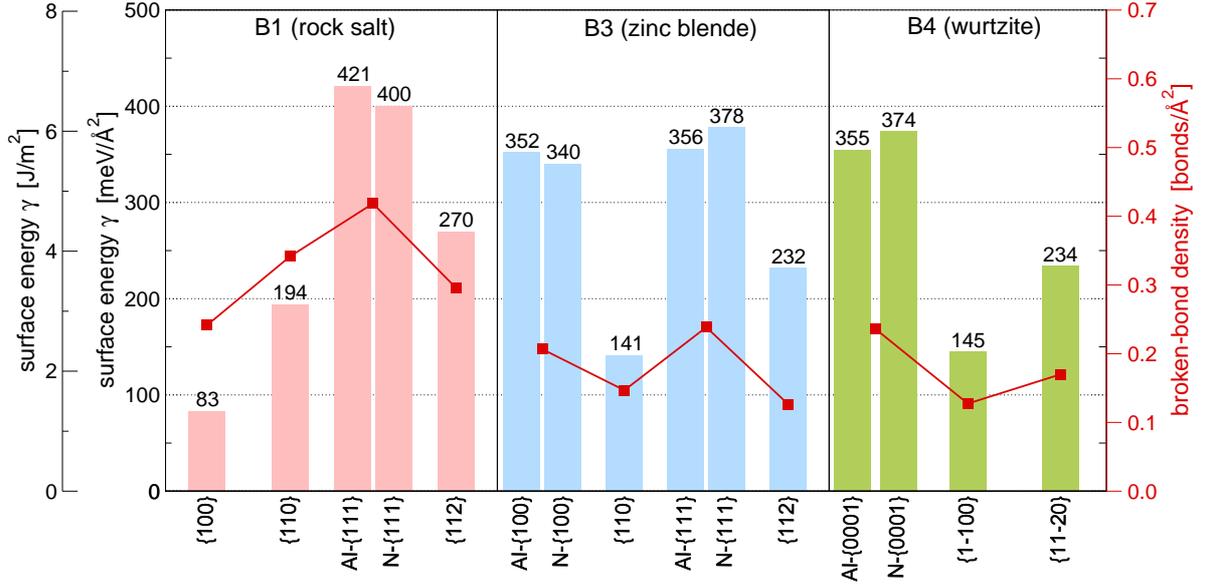}
  \caption{Surface energies for low-index planes of AlN allotropes. Red curves represent calculated density of broken bonds as discussed in the text.}\label{fig:gamma}
\end{figure}

The resulting surface energies are shown in Fig.~\ref{fig:gamma}. The highest values of $\gamma$ (between $340$--$420\uu{meV/\AA}^2$ are for all structures obtained for mono-specie surfaces. Additionally, the difference between surfaces terminated by Al or N atom are $\approx20\uu{meV}$ or less. This means that in the cases when only the orientation rather than the exact structure of the surface is known, one can used the average value of Al- and N-terminated surfaces as a reasonable estimate without changing e.g., the energetic order of individual facets. The surface energy of facets formed by both atomic species (stoichiometric facets) is by at least $100\uu{meV/\AA}^2$ or more lower than the mono-atomic ones. The lowest surface energies of individual structures posses $\{100\}$ planes for B1, $\{110\}$ planes for B3 and $\{1\bar100\}$ planes for B4.

The observed trends can be rationalised by a simple nearest-neighbour broken-bond model \cite{Kozeschnik2007,Becker1938}. If one assumes that the main contribution to the surface energy is the ``penalty'' related to breaking bonds that cross the actual surface, then $\gamma$ should be proportional to $N_{\rm bonds}/A$, where $N_{\rm bonds}$ is the number of broken bonds per surface area $A$ of the primitive cell. This density of broken bonds is also plotted in Fig.~\ref{fig:gamma}. It captures the surface energy trends surprisingly well, the only deficiencies being the mutual relations of the cubic $\{110\}$ and $\{112\}$ facets. Since no plane can cut more then 3 bonds of the octahedrally coordinated sites in the B1 structure (which is the case for $\{112\}$ plane), and since the surface area per atom increases for planes with higher Miller indices, the nearest-neighbour broken-bond model suggests that the surface energy of all those planes is less than $\gamma_{\{112\}}=270\uu{meV/\AA}^2$. The situation is slightly more complex for the octahedrally coordinates B3 and B4 structures. If always a pair of mirror interface is considered (as in the case of this study where e.g., $\gamma_{\{0001\}}^{\rm B4}=\frac12(\gamma_{(0001)}^{\rm B4}+\gamma_{(000\bar1)}^{\rm B4})$), then the maximum number of broken bonds per surface atom is 2. The only possibility how to get a surface with atoms having only 1 broken bond is that 2 out of 4 bonds lie in the surface. In all other cases, the surface atoms have 2 broken bonds per atom on average. Out of those, the B3-$\{111\}$ and B4-$\{0001\}$ planes have the smallest area per atom, thus justifying why $\gamma_{\{111\}}^{\rm B3}$ and $\gamma_{\{0001\}}^{\rm B4}$ are the highest surface energies for those structures.

\section{Conclusions}

We used the quantum mechanical calculations to obtain surface energies of low-index surface facets of rock-salt B1, zinc-blende B3, and wurtzite B4 allotropes of AlN. No surface reconstruction was taken into account. The highest surface energies were obtained for mono-specie planes i.e., $\gamma_{\{111\}}^{\rm B1}=410\uu{meV/\AA}^2$, $\gamma_{\{100\}}^{\rm B3}=346\uu{meV/\AA}^2$, $\gamma_{\{111\}}^{\rm B3}=360\uu{meV/\AA}^2$, and $\gamma_{\{0001\}}^{\rm B4}=365\uu{meV/\AA}^2$. The observed trends could be rationalised by the nearest-neighbour broken-bond model: the higher the areal density of broken bonds, the higher the surface energy.

\corr{
\section*{Acknowledgements}

Financial support by Christian Doppler Research Association as well as by the START Program (Y371) of the Austrian Science Fund (FWF) is greatly acknowledged.}

\appendix

\begin{table}[!p]
\section{Structures}\label{appendix}

\centering
\caption{Lattice vectors and fractional coordinates of atoms for the primitive cells used for construction of $\{100\}$, $\{110\}$, $\{111\}$, and $\{112\}$ surfaces of the B1 structure. The N-terminated $\{111\}$ slab is obtained by exchanging Al and N lattice sites.}
\begin{tabular}{c|rrr|rrr|rrr|rrr}
                  & \multicolumn{3}{c|}{$\{100\}$} & \multicolumn{3}{c|}{$\{110\}$} & \multicolumn{3}{c|}{$\{111\}$}  & \multicolumn{3}{c}{$\{112\}$}    \\ \hline
  $\mathbf{a}_1$  & $1/2$ & $-1/2$ & $0$           & $\sqrt2/2$ & $0$ & $0$         & $\sqrt2/4$ & $-\sqrt6/4$ & $0$  & $\sqrt3$ & $0$ & $0$     \\
  $\mathbf{a}_2$  & $1/2$ & $1/2$ & $0$            & $0$ & $1$ & $0$                & $\sqrt2/4$ & $\sqrt6/4$ & $0$   & $0$ & $\sqrt2/2$ & $0$    \\
  $\mathbf{a}_3$  & $0$ & $0$ & $1$                & $0$ & $0$ & $\sqrt2/2$               & $0$ & $0$ & $\sqrt3$            & $0$ & $0$ & $\sqrt6/2$              \\ \hline
  Al              & $0$ & $0$  & $0$               & $0$ & $0$ & $0$                & $0$ & $0$ & $0$                 & $0$ & $0$ & $0$                   \\
  Al              & $1/2$ & $1/2$ & $1/2$          & $1/2$ & $1/2$ & $1/2$          & $1/3$ & $1/3$ & $1/3$           & $2/3$ & $1/2$ & $1/6$             \\
  Al              &   &  &                         &   &   &                        & $2/3$ & $2/3$ & $2/3$           & $1/3$ & $0$ & $1/3$             \\
  Al              &   &  &                         &   &   &                        &    &  &                         & $0$ & $1/2$ & $1/2$             \\
  Al              &   &  &                         &   &   &                        &    &  &                         & $2/3$ & $0$ & $2/3$             \\
  Al              &   &  &                         &   &   &                        &    &  &                         & $1/3$ & $1/2$ & $5/6$             \\
  N               & $1/2$ & $1/2$ & $0$            & $0$ & $1/2$ & $0$              & $0$ & $0$ & $1/2$               & $1/2$ & $0$ & $0$                 \\
  N               & $0$ & $0$ & $1/2$              & $1/2$ & $0$ & $1/2$            & $1/3$ & $1/3$ & $5/6$           & $1/6$ & $1/2$ & $1/6$             \\
  N               &   &  &                         &   &   &                        & $2/3$ & $2/3$ & $1/6$           & $5/6$ & $0$ & $1/3$             \\
  N               &   &  &                         &   &   &                        &    &  &                         & $1/2$ & $1/2$ & $1/2$             \\
  N               &   &  &                         &   &   &                        &    &  &                         & $1/6$ & $0$ & $2/3$             \\
  N               &   &  &                         &   &   &                        &    &  &                         & $5/6$ & $1/2$ & $5/6$             \\
\end{tabular}\bigskip

\caption{Lattice vectors and fractional coordinates of atoms for the primitive cells used for construction of $\{100\}$, $\{110\}$, $\{111\}$, and $\{112\}$ surfaces of the B3 structure. The N-terminated $\{100\}$ and $\{111\}$ slabs are obtained by exchanging Al and N lattice sites.}
\begin{tabular}{c|rrr|rrr|rrr|rrr}
                  & \multicolumn{3}{c|}{$\{100\}$} & \multicolumn{3}{c|}{$\{110\}$} & \multicolumn{3}{c|}{$\{111\}$}  & \multicolumn{3}{c}{$\{112\}$}    \\ \hline
  $\mathbf{a}_1$  & $1/2$ & $-1/2$ & $0$           & $\sqrt2/2$ & $0$ & $0$         & $\sqrt2/4$ & $-\sqrt6/4$ & $0$  & $\sqrt3$ & $0$ & $0$     \\
  $\mathbf{a}_2$  & $1/2$ & $1/2$ & $0$            & $0$ & $1$ & $0$                & $\sqrt2/4$ & $\sqrt6/4$ & $0$   & $0$ & $\sqrt2/2$ & $0$    \\
  $\mathbf{a}_3$  & $0$ & $0$ & $1$                & $0$ & $0$ & $\sqrt2/2$               & $0$ & $0$ & $\sqrt3$            & $0$ & $0$ & $\sqrt6/2$              \\ \hline
  Al              & $0$ & $0$  & $0$               & $0$ & $0$ & $0$                & $0$ & $0$ & $0$                 & $0$ & $0$ & $0$                   \\
  Al              & $1/2$ & $1/2$ & $1/2$          & $1/2$ & $1/2$ & $1/2$          & $1/3$ & $1/3$ & $1/3$           & $2/3$ & $1/2$ & $1/6$             \\
  Al              &   &   &                        &   &   &                        & $2/3$ & $2/3$ & $2/3$           & $1/3$ & $0$ & $1/3$             \\
  Al              &  &  &                          &   &   &                        &    &  &                         & $0$ & $1/2$ & $1/2$             \\
  Al              &   &  &                         &   &   &                        &    &  &                         & $2/3$ & $0$ & $2/3$             \\
  Al              &   &  &                         &   &   &                        &    &  &                         & $1/3$ & $1/2$ & $5/6$             \\
  N               & $1/2$ & $0$ & $3/4$            & $0$ & $1/4$ & $1/2$            & $0$ & $0$ & $3/4$               & $1/4$ & $0$ & $0$                 \\
  N               & $0$ & $1/2$ & $1/4$            & $1/2$ & $3/4$ & $0$            & $1/3$ & $1/3$ & $1/12$          & $11/12$ & $1/2$ & $1/6$             \\
  N               &   &   &                        &   &   &                        & $2/3$ & $2/3$ & $5/12$          & $7/12$ & $0$ & $1/3$             \\
  N               &   &   &                        &   &   &                        &    &  &                         & $1/4$ & $1/2$ & $1/2$             \\
  N               &   &  &                         &   &   &                        &    &  &                         & $11/12$ & $0$ & $2/3$             \\
  N               &   &  &                         &   &   &                        &    &  &                         & $7/12$ & $1/2$ & $5/6$             \\
\end{tabular}
\end{table}

\begin{table}[!ht]
\centering
\caption{Lattice vectors and fractional coordinates of atoms for the primitive cells used for construction of $\{0001\}$, $\{1\bar100\}$, and $\{11\bar20\}$ surfaces of the B4 structure. The N-terminated $\{0001\}$ slab is obtained by exchanging Al and N lattice sites. $u$ is the internal parameter describing shift of N atoms above Al sites ($u_{\rm AlN}=0.382$). $c$ and $a$ are the lattice parameters.}
\begin{tabular}{c|rrr|rrr|rrr}
                  & \multicolumn{3}{c|}{$\{0001\}$} & \multicolumn{3}{c|}{$\{1\bar100\}$} & \multicolumn{3}{c}{$\{11\bar20\}$}  \\ \hline
  $\mathbf{a}_1$  & $1/2$ & $-\sqrt3/2$ & $0$      & $1$ & $0$ & $0$                & $3/2$ & $0$ & $0$  \\
  $\mathbf{a}_2$  & $1/2$ & $\sqrt3/2$ & $0$       & $0$ & $c/a$ & $0$              & $0$ & $c/a$ & $0$   \\
  $\mathbf{a}_3$  & $0$ & $0$ & $c/a$              & $0$ & $0$ & $\sqrt3$           & $0$ & $0$ & $\sqrt3/2$         \\ \hline
  Al              & $1/3$ & $2/3$  & $0$           & $0$ & $0$ & $0$                & $0$ & $0$ & $0$                    \\
  Al              & $2/3$ & $1/3$ & $1/2$          & $1/2$ & $1/2$ & $1/6$          & $1/3$ & $1/2$ & $0$               \\
  Al              &   &  &                         & $1/2$ & $0$ & $1/2$            & $1/2$ & $0$ & $1/2$                       \\
  Al              &   &  &                         & $0$ & $1/2$ & $2/3$            & $5/6$ & $1/2$ & $1/2$                    \\
  N               & $1/3$ & $2/3$ & $u$            & $0$ & $u$ & $0$                & $0$ & $u$ & $0$                 \\
  N               & $2/3$ & $1/3$ & $1/2+u$        & $1/2$ & $1/2+u$ & $1/6$        & $1/3$ & $1/2+u$ & $0$           \\
  N               &   &  &                         & $1/2$ & $u$ & $1/2$            & $1/2$ & $u$ & $1/2$                      \\
  N               &   &  &                         & $0$ & $1/2+u$ & $2/3$          & $5/6$ & $1/2+u$ & $1/2$                \\
\end{tabular}
\end{table}

%

\end{document}